\documentclass[useAMS,usenatbib]{mn2e} \usepackage{latexsym}
\usepackage{times} \usepackage{graphics}
\usepackage[german,english,american]{babel} \newcommand{\apj}{ApJ}
\newcommand{\apjl}{ApJL} 
\newcommand{\aap}{A\&A} \newcommand{\mnras}{MNRAS}

\newcommand{\pasp}{PASP}
\newcommand{\sci}{Science}
 
\newcommand{\arcdeg}{^{\circ}}
\def\lesssim{\mathrel{\hbox{\rlap{\hbox{\lower4pt\hbox{$\sim$}}}\hbox{$<$}}}}
\def\gtrsim{\mathrel{\hbox{\rlap{\hbox{\lower4pt\hbox{$\sim$}}}\hbox{$>$}}}}

\title[Constraints on beaming]{Constraints on relativistic beaming
  from estimators of the unbeamed flux}

\author[Heinz \& Merloni]{S.~Heinz$^{1,2}$ \& A.~Merloni$^{3}$\\$^{1}$
  Center for Space research, MIT, 77 Mass. Ave., Cambridge, MA
  02139\\$^{2}$ Chandra Fellow\\$^{3}$ Max-Planck-Institute for
  Astrophysics, Karl-Schwarzschild-Str.~1, 85743 Garching, Germany}
\begin{document}
\date{1 October 2004} \maketitle

\begin{abstract}
We review the statistical properties of relativistic Doppler boosting
relevant for studies of relativistic jets from compact objects based
on radio--X-ray(--mass) correlations, such as that found in black-hole
X-ray binaries in the low/hard state, or the ``fundamental plane'' of
Merloni, Heinz, \& DiMatteo.  We show that the presence of only
moderate scatter in such relations does not necessarily imply low
Lorentz factors of the jets producing the radio emission in the
samples under consideration.  Applying Doppler beaming statistics to a
large sample of XRBs and AGN, we derive a limit on the width of the
Lorentz factor distribution of black holes with relativistic jets: If
the X-rays are unbeamed (e.g., if they originate in the accretion disk
or in the slower, innermost part of the jet), the width of the
$\beta\Gamma$ distribution should be about one order of magnitude or
less.  If the scatter about the ``fundamental plane'' is entirely
dominated by relativistic beaming, a lower limit on the mean Lorentz
factor $\langle\beta\Gamma\rangle > 5$ can be derived.  On the other
hand, if the X-rays are boosted by the same factor as the radio
emission, we show that the observed scatter cannot be reasonably
explained by Doppler boosting alone.
\end{abstract}
\begin{keywords}
\end{keywords}
\section{Introduction}
\label{sec:introduction}
Astrophysical jets from black holes in active galactic Nuclei (AGNs)
and X-ray binaries (XRBs) are known to propagate with velocities close
to the speed of light.  Evidence for relativistic bulk motion stems
from variability and compactness limits in compact cores
\citep[][]{jones:74}, from superluminal motions of
knots\citep[][]{whitney:71,cohen:71}, and from doppler boostes
jet-to-counter jet flux ratios \citep[e.g.][]{perley:82}. While proper
motion measurements typically only constrain pattern speeds, the body
of evidence points towards a range of mildly to ultra-relativistic jet
speeds.  As a result, the observed flux densities from these jets can
be affected by strong Doppler boosting, which must be corrected for if
one wants to infer the intrinsic luminosity of the jet, emitted in the
rest frame of the plasma, which is important when calculating the
physical parameters of the jet.

Recently, a correlation between the radio emission from steady jets
and the hard (2-10 keV) X-ray emission has been found in black hole
XRBs \citep{corbel:03,gallo:03}.  This relation implies that in a
given XRB in the low/hard state \citep[see][for a review of the state
classification in XRBs]{mcclintock:04}, the radio flux is proportional
to the X-ray flux to the 0.7th power.  This trend has been observed
for different sources.  The normalisation of the relation differs only
by factors of a few for different XRBs.  Since the radio emission
stems from the jet, it will be affected by Doppler boosting and its
brightness will differ strongly for different viewing angles.  In the
standard scenario, the X-rays come from the disk (or alternatively the
inner, slow moving part of the jet) and are thus not significantly
beamed.  The lack of strong scatter in the radio-X-ray relation has
been interpreted as an indication that Doppler boosting is weak in the
steady jets of XRBs in the low/hard state and that they must therefore
move with only mildly relativistic speeds \citep{gallo:03}.

Furthermore, \cite{merloni:03} and \citep{falcke:04} have found a
strong correlation between radio luminosity, X-ray luminosity, and
black hole mass for samples of black holes spanning a wide range in
black hole masses and accretion rates (called the ``fundamental plane
of black hole activity'', FP for short).  Again, this
relation shows a certain amount of intrinsic scatter, part of which
might be contributed by Doppler boosting.  Thus, studying this scatter
can provide constraints on the presence or absence of Doppler boosting
in the jets that produce the radio emission.

The nature of Doppler boosting has been studied exhaustively in the
literature, specifically in regard to its statistical effects on
samples of objects emitting beamed radiation
\citep{orr:82,urry:84,urry:91,urry:95,morganti:95,lister:97,lister:03}.
Typically, one deals with a sample of radio sources that have been
selected in the radio band and consider the effects of Doppler
boosting on their luminosity function and on possible selection
effects, not knowing what the unboosted flux of any particular source
in the sample is.

What makes the situation considered in this paper different is that we
actually {\em have} an unbiased estimator of the unbeamed radiation:
in the case of XRBs it is the X-ray luminosity, in the more general
case of black holes of all masses it is the FP relation that links
radio luminosity, X-ray luminosity, and black hole mass.  It is
therefore worth considering the statistical properties of relativistic
Doppler boosting under those conditions.

In \S\ref{sec:boosting} we will review the basic properties of Doppler
boosting and define the statistical integrals necessary for the
remainder of the paper.  In \S\ref{sec:binaries} we will apply these
results to individual pairs of XRBs and argue that the observed
moderate amount of scatter in the XRB radio-X-ray relation alone
cannot be used to argue for low jet velocities.  In \S\ref{sec:agn} we
apply the same method to the FP sample to derive constraints on the
Lorentz factor distribution of the source in the sample.  Section
\ref{sec:conclusion} presents our conclusions.

\section{The beaming probability distribution}
\label{sec:boosting}
In the following we will consider radio emission from two-sided jets.
We will assume that the approaching jet is identical to the receding
jet.  We will further assume that the spectrum emitted by the jet is a
powerlaw with index $\alpha_{\rm r}$ such that the jet flux is
$F_{\nu} \propto \nu^{-\alpha_{\rm r}}$.  We will use a fiducial value
of $\alpha_{\rm r}=0$, appropriate for the cores of jets observed in
AGNs and XRBs, which show a roughly flat spectrum emitted from a {\em
continuous} jet. For a review on jet properties and relativistic
beaming, see, e.g., \cite{begelman:84}.

We are interested in situations where we have an independent estimator
of the relative radio flux of different sources in the sample from
observables like the X-ray flux, the distance, and the black hole
mass, such as were proposed by
\cite{corbel:03,gallo:03,merloni:03,falcke:04}.  Furthermore, we are
interested in situations where the set of independent measurements is
drawn from a sample of sources that are not selected in the spectral
band where beaming is important and can thus be assumed to be oriented
randomly.  That is, the orientation of the approaching jet is random
on a hemisphere of $2\pi$ steradian.  This implies that $\cos{\theta}$
is randomly distributed between 0 and 1, where $\theta$ is the angle
between the line of sight and the approaching jet.  It does {\em not}
imply that $\theta$ is randomly distributed between 0 and $\pi/2$.

For a given jet Lorentz factor $\Gamma$ and four-velocity $\beta\Gamma
= \sqrt{\Gamma^2 - 1}$, the relativistic Doppler boosting formula for
the observed flux $F_{\nu}$ relative to the flux emitted in the rest
frame of the plasma $F_{\nu,\rm jet}$ is:
\begin{equation}
  F_{\nu} = \frac{F_{\nu,\rm jet}}{{\Gamma}^{k+\alpha_{\rm
    r}}}\left[\frac{1}{\left(1 +
    \beta\cos{\theta}\right)^{k+\alpha_{\rm r}}} + \frac{1}{\left(1 -
    \beta\cos{\theta}\right)^{k+\alpha_{\rm r}}}\right]
    \label{eq:beaming}
\end{equation}
where $k$ varies from 2 for continuous jets to 3 for discrete
ejections \citep[e.g.][]{urry:95}.  Since we are considering steady,
quasi-continuous jets, we will take $k=2$ as our fiducial value.

Since the sources are randomly oriented, the fraction $P(>\theta)$ of
sources with line of sight angle larger than $\theta$ is simply
\begin{equation}
  P(>\theta) = \int_{0}^{\theta} d\theta \sin{\theta} = \cos{\theta}
\end{equation}
(note that $0\arcdeg \leq \theta \leq 90\arcdeg$) and
eq.~(\ref{eq:beaming}) becomes
\begin{equation}
  F_{\nu}(P,\Gamma) = \frac{F_{\rm jet}(\nu)}
  {\Gamma^{k+\alpha_{\rm r}}}\left[\frac{1}{\left(1 + \beta
  P\right)^{k+\alpha_{\rm r}}} + \frac{1}{\left(1 - \beta
  P\right)^{k+\alpha_{\rm r}}}\right]
  \label{eq:flux}
\end{equation}
$F(P,\Gamma)$ is monotonic in $P$ for $0 \leq P \leq 1$.  For a given
$\Gamma$, $P$ is the probability to observe a source at boosted flux
lower than $F_{\nu}(P)$.

\begin{figure}
\begin{center}
\resizebox{0.95\columnwidth}{!}{\includegraphics{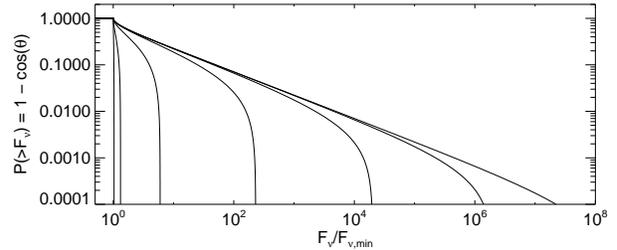}}
\end{center}
\caption{Probability for Doppler boosted flux to lie above a given
  value (relative to the maximally de-boosted flux $F_{\rm min}$ at
  $90^{\arcdeg}$ viewing angle).  Curves are for increasing 4-velocity
  $\beta\Gamma$ from $\beta\Gamma=0.1$ (leftmost) to $\beta\Gamma=100$
  (rightmost), in logarithmic intervals increasing by factors of
  $\sqrt{10}$.
  \label{fig:probdist}}
\end{figure}

Assuming the fiducial values of $\alpha_{\rm r}=0$ and $k=2$, we can
invert eq.~(\ref{eq:flux}) to find the cumulative probability of
observing a source at a flux lower than $F_{\nu}$ (plotted in
Fig.~\ref{fig:probdist}):
\begin{equation}
  P(<F_{\nu})=\frac{1}{\beta}\sqrt{1 + \frac{1}{F\Gamma^2} -
  \sqrt{\left(1 + \frac{1}{F\Gamma^2}\right)^2 + \frac{2}{F\Gamma^2} -
  1}}
  \label{eq:prob}
\end{equation}
and, conversely, $P(>F_{\nu}) = 1 - P(<F_{\nu})$.

It is clear from eq.(\ref{eq:prob}) that, in a randomly oriented
sample of jets with identical $\Gamma$, most of the sources fall into
a relatively narrow flux range: Using $0 \leq \beta \leq 1$, we can
see that 50\% of the sources fall within the range
\begin{eqnarray}
  \frac{2}{\Gamma^{k+\alpha_{\rm r}}} \leq & F_{\nu} & \leq
  \frac{1}{\Gamma^{k+\alpha_{\rm r}}}\left(2^{k+\alpha_{\rm r}} +
  \left(\frac{2}{3}\right)^{k+\alpha_{\rm r}}\right)
\end{eqnarray}
For the fiducial parameters, these two limits fall within a factor of
$\leq 2.2$.  Thus, {\em independent} of the actual Lorentz factor, the
fluxes of 50\% of the sources in a randomly oriented sample of flat
spectrum jets with identical $\Gamma$ fall within a factor of $\leq
2.2$.  The remaining sources are distributed in a tail to larger
observed fluxes, cutting off at the maximum flux, $F_{\nu} \leq
\Gamma^2\left(2 + 2\beta^2\right)$ (see Fig.~\ref{fig:probdist}).  The
curves for $\Gamma=10$ and $\Gamma=100$ differ only below $P(>F_{\nu})
< 3$\%, i.e., for 3 out of 100 sources.

Well below this cutoff, the probability distribution is very similar
for different $\Gamma$, but shifted to lower fluxes (i.e., deboosted)
by a factor of $\zeta_{0} \equiv 1/\Gamma^{k+\alpha_{\rm r}}$.  Thus,
measuring the width of the flux distribution (FWHM) of randomly
oriented sources {\em with identical $\Gamma$} is not sufficient to
determine $\Gamma$ if the width is larger than about a factor of 2.2.
A proper determination would require sampling the cutoff.  For a
measured width of $\delta \equiv F_{\rm max}/F_{\rm min} \gg 2.2$, to
be able to say that the upper limit corresponds to the cutoff would
require a total number of sources well in excess of
\begin{equation}
  N(\delta)=\left[1 - \sqrt{\frac{2\delta + 1 - \sqrt{8\delta +
  1}}{2\delta}}\right]^{-1}
  \label{eq:number}
\end{equation}
where we used eq.~(\ref{eq:flux}) and $\alpha=0$ and $k=2$.

However, it is rather unlikely that the bulk Lorentz factors of all
the sources are identical.  Instead, $\beta\Gamma$ will follow some
distribution $f(\beta\Gamma)$ around a mean $\beta\Gamma_{\rm
mean}=\langle\beta\Gamma\rangle$.  Because of the strong dependence of
the shift $\zeta_{0}$ of the flux distribution on $\Gamma$ and because
the flux distribution for a given $\Gamma$ is strongly concentrated
around the minimum value, the spread $\delta$ in the flux distribution
{\em of a randomly oriented sample} is typically dominated by the
spread in $\Gamma$, not by viewing angle effects.

Thus, for flux distributions significantly wider than $\delta \sim
2.2$, we cannot determine the maximum or mean $\beta\Gamma$ simply by
measuring the width of the flux distribution, assuming an inherently
uniform flux, or by imposing a universal radio-X-ray relation
\citep{gallo:03,merloni:03,falcke:04} and measuring the spread against
this relation.  Only if the distribution contains a total number of
sources well in excess of the value of eq.~(\ref{eq:number}) and if
the upper cutoff of the flux distribution is well sampled can one
derive an upper limit on $\beta\Gamma$.  Otherwise, the only
conclusion that can be reached from a relatively narrow distribution
in fluxes around a radio-X-ray relation is that {\em the spread} in
$\Gamma$ around $\beta\Gamma_{\rm mean}$ is small.

\section{Constraints for a single pair of radio jets}
\label{sec:binaries}

For situations where a tight relation between the beamed radio jet
emission with some unbeamed observables (e.g., the X-ray flux) is
observed over a large range in the secondary observables but for a
small number of sources, one can derive constraints on the Lorentz
factors of individual pairs of sources from the difference in
normalisation of the observed relation, assuming that it reflects only
differences in orientation and Lorentz factor.

One such example is the XRB radio-X-ray relation
\citep{corbel:03,gallo:03}, where the number of sources contributing
is rather small --- between 2 and 4 on the low luminosity end, where
the relation holds most firmly.  The two most significant sources in
the sample are V404 Cyg and GX339-4.  Following \cite{gallo:03}, the
radio flux in V404 is a factor of about 2.5 to 5 larger than that of
GX339-4 for the same X-ray flux.  Allowing for some uncertainty in the
mass of the black hole in GX339-4 \citep{hynes:04} and of the
distances to GX339-4 and V404 \citep{hynes:04,jonker:04}, the rough
confidence limits on this ratio fall between 1.5 and 5.  We can then
ask what constraints on beaming can be derived from this observation.

We assume that, at the same X-ray luminosity, both sources have the
same comoving (i.e., unbeamed) radio luminosity, i.e., they fall on
the same X-ray-radio relation when corrected for beaming.  In other
words, we assume that the X-rays are not affected by beaming (see
\S\ref{sec:agn} for more discussion of this assumption).  If the
jets have Lorentz factors of $\Gamma_{404}$ and $\Gamma_{339}$, the
probability that the observed radio flux from V404 is larger than that
of GX339 by a factor $\delta$ is
\begin{equation}
  P(F_{404}>\delta F_{339}) = 1 - \int_{0}^{1} dp P\left(\delta
  F_{\nu}(p,\Gamma_{339}), \Gamma_{404}\right)
\end{equation}
where $F_{\nu}(p,\Gamma)$ follows eq.~(\ref{eq:flux}) and
$P(F,\Gamma)$ is taken from eq.~(\ref{eq:prob}).

\begin{figure}
\begin{center}
\resizebox{\columnwidth}{!}{\includegraphics{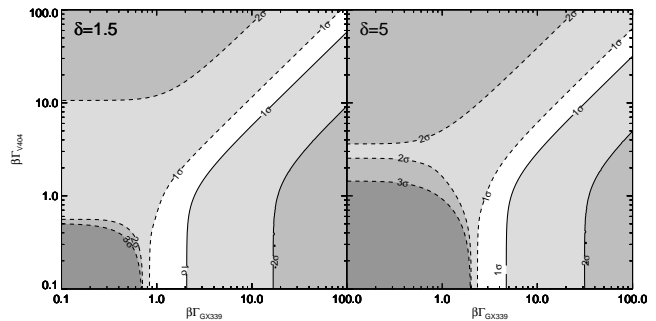}}
\end{center}
\caption{Chisquare maps of the jet Lorentz factors of GX339-4 (bottom
  axis) and V404 Cyg (left axis), given the observed difference in
  radio luminosity $1.5 \lesssim \delta \lesssim 5$ (left and right
  panel respectively).  Shown are the contours outside of which the
  probability of observing a flux ratio larger or smaller then
  $\delta$ are 68\%, 95\%, 99.5\% (i.e., 1,2, and 3 sigma).
  \label{fig:scatter1}}
\end{figure}

Fig.~\ref{fig:scatter1} shows the one-, two-, and three-sigma contours
on $\beta\Gamma$ of both jets for the range in normalisation offsets
allowed by the observations \citep{gallo:03} $1.5 \lesssim
F_{404}/F_{339} \lesssim 5$.  The fact that the ratio of
$F_{404}/F_{339}$ is close to unity implies that the Lorentz factors
of both sources fall within roughly a factor of 2 and that the jet in
GX339-4 likely has a higher Lorentz factor than that of V404.  The
possible presence of larger uncertainties in black hole mass and
distance to both objects that are unaccounted for in our estimate of
$\delta$ imply that the confidence contours in Fig.~\ref{fig:scatter1}
will be widened and the constraints on
$\beta\Gamma_{404}/\beta\Gamma_{339}$ will be less stringent, thus
allowing the $\Gamma$ of both objects to be more different than
otherwise implied.

While it is not possible to extend this graphical analysis to more
than 2 sources, the formalism can easily be adapted to N sources, in
which case the confidence contours turn into N-1 dimensional
hyper-surfaces in an N dimensional $\log{\beta\Gamma_{n}}$ space.
Asymptotically (at large $\beta\Gamma_{n}$), the surfaces will
describe hyper-cylinders around an axis parallel to the diagonal
vector $(1,1,...,1)$, shifted along each axis by the square root of
the flux ratio of the reference source relative to source $n$.  For
large $N$, this distribution of shifts is then a representation of the
distribution of $\Gamma_{n}$.

The formal conclusion we reach from this analysis is that the relative
similarity in the normalisation of the radio-X-ray relations for
GX339-4 and V404 Cyg does not imply that the Lorentz factors of both
jets are small, but rather that they are similar.  From the
constraints on the scatter about the radio-X-ray relation, we cannot
put any upper limit on $\Gamma$ of either source.  However, because
for large $\Gamma$, the observed radiation is severely de-boosted,
other physical limitations can provide such limits.  E.g., at very
large $\Gamma$, the implied kinetic power would vastly exceed any
reasonable limits \citep{fender:04}.  Also, radio timing constraints
from Cyg X-1 indicate that its jet is only moderately relativistic
\citep{gleissner:04}.

\section{The spread in the fundamental plane}
\label{sec:agn}
We will now use the scatter observed in the radio--X-ray--mass
``fundamental plane'' (FP) correlation found by \cite{merloni:03} and
\cite{falcke:04} to constrain the Lorentz factor distribution of the
jets in the sample.  These limits will be based on the assumption that
the orientation of the sources is random and that the scatter in the
distribution is at least partly due to relativistic beaming. Clearly,
other sources of scatter will enter (e.g., uncertainty in black hole
mass, spin, variations in $\alpha_{\rm r}$), so the observed scatter
cannot be {\em solely} due to relativistic boosting.  This implies
that any constraints derived here will be upper limits.  We will show
that the observed scatter can only be used to constrain the width of
the Lorentz factor distribution.

\subsection{Unbeamed X-rays}
If the X-ray emission of the sources in the sample stems from the
accretion disk, the X-rays will not be affected by relativistic
beaming.  It should be noted that the disk X-ray emission can still be
anisotropic simply due to the nature of the accretion flow
\citep[e.g.,][]{shakura:73,beloborodov:99}, however, the scatter
produced by the differences in viewing angle is a relatively mild
effect and small compared to the scatter due to boosting, and we will
neglect this effect in the following.  We can estimate the radio
Doppler boosting factor from eq.~(\ref{eq:flux}) using $k=2$ and
$\alpha_{\rm r}=0$.  We can then relate this expression to the scatter
about the FP,
\begin{equation}
  \delta = F_{\rm r}/(10^{7.33}F_{\rm x}^{0.6}M^{0.78})
	\label{eq:delta}
\end{equation}

\begin{figure}
\begin{center}
\resizebox{\columnwidth}{!}{\includegraphics{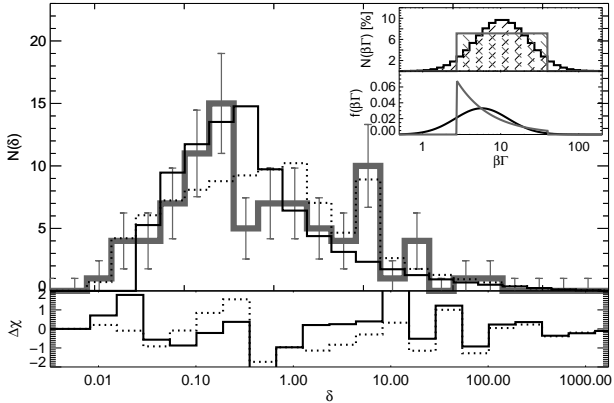}}
\end{center}
\caption{Upper panel: {\em thick grey curve}: Histogram of the scatter
  of all individual sources of the FP sample (excluding upper limits
  and averaging over the different data sets for different XRBs); {\em
  solid black curve}: fit of log-normal distribution with
  $\beta\Gamma_{\rm mean}=10.5$, $\sigma=0.78$; {\em dotted black
  curve}: fit of log-flat $\beta\Gamma$ distribution with
  $\beta\Gamma_{\rm mean}=10.5$, $\sigma=1.34$. Lower panel:
  residuals. Insert: {\em Upper panel}: histogram of
  $\beta\Gamma$-distributions corresponding to fits in
  Fig.~\ref{fig:histogram} ({\em black}: log-normal, {\em grey}:
  log-flat).  {\em Lower panel}: corresponding $\beta\Gamma$
  distributions.
  \label{fig:histogram}}
\end{figure}

Since we have no information about the distribution of $\Gamma$, we
will take two simple functional forms as templates.  First, we will
use a log--normal distribution of the form (see
Fig.~\ref{fig:histogram}):
\begin{equation}
  f(\beta\Gamma)=\frac{N\exp{\left[-\left(\log{\beta\Gamma/\beta\Gamma_{\rm
  mean}}\right)^2/\left(2\sigma^2\right)\right]}}
  {\beta\Gamma\sigma\sqrt{2\pi}}
  \label{eq:lognormal}
\end{equation}
Since the un-beamed normalization of the radio flux is unknown (the
mean in the FP distribution corresponds to an average over all angles
and $\Gamma$'s), we have to allow for an arbitrary re-normalisation of
the flux $\delta_0$.  We can then produce a histogram of the scatter
$\delta$ of all the sources in the FP relation.  This is shown in
Fig.~\ref{fig:histogram}.  We have used Poisson errors for the
histogram bins.  Also shown is a fit of a log-normal distribution in
$\beta\Gamma$ to this histogram (fit parameters: $\beta\Gamma_{\rm
mean}=7$, $\sigma=0.78$, $\delta_0=0.74$), which can
reproduce the range and shape of the scatter distribution rather well.

Fitting a log-flat distribution of the form
$f(\beta\Gamma)=N/\left(\ln{(\sigma^2)}\beta\Gamma\right)$
for $\frac{\beta\Gamma_{\rm mean}}{\sigma} \leq
\beta\Gamma \leq \sigma\beta\Gamma_{\rm mean}$ and
$f(\beta\Gamma)=0$ elsewhere, provides a marginally better fit, which
can be understood by the fact that it is a decent approximation to
eq.~(\ref{eq:lognormal}) to lowest order.  This shows that we cannot
constrain the shape of the $\beta\Gamma$ distribution very well.  For
the purpose of this letter, we shall limit ourselves to constraining
the width of this distribution.  A better determination of the shape
of the distribution will only be possible when a larger, more
carefully selected sample is available.

\begin{figure}
\begin{center}
\resizebox{\columnwidth}{!}{\includegraphics{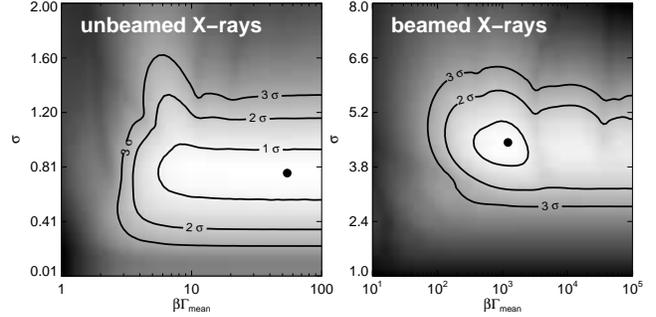}}
\end{center}
\caption{Chisquare maps of the two interesting parameters
   $\beta\Gamma_{\rm mean}$ and $\sigma$, fitting a
   log--normal distribution in $\Gamma$ to the $\delta$-histogram in
   Fig.~\ref{fig:histogram}. The curves show the formal 1, 2, and 3
   sigma confidence contours. Left panel: unbeamed X-rays, right
   panel: beamed X-rays.\label{fig:chisquare}}
\end{figure}

In \S\ref{sec:introduction} we argued that the width of the scatter
distribution about a radio--X-ray(--mass) relation can only be used to
constrain the width $\sigma$ of the distribution, not
$\beta\Gamma_{\rm mean}$ itself.  To demonstrate this point
quantitatively, Fig.\ref{fig:chisquare} shows the chi-square
distribution of the two interesting parameters $\beta\Gamma_{\rm
mean}$ and $\sigma$ (marginalising over the unknown
radio flux normalisation $\zeta_0$ of the underlying, unbeamed FP
relation) of the assumed lognormal distribution in $\beta\Gamma$ used
to fit the $\delta$ histogram in Fig.~\ref{fig:histogram}.  The 1, 2,
and 3 sigma confidence contours show that $\sigma$ is
constrained much better than $\beta\Gamma_{\rm mean}$.  In fact, the
fit only provides a {\em lower limit} on $\beta\Gamma_{\rm mean}$,
similar to the result in Fig.~\ref{fig:scatter1}.  However, since
other sources of scatter will render all measurements derived from the
scatter about the FP upper limits, we cannot make any statements about
the mean $\beta\Gamma_{\rm mean}$ in the sample, while we can safely
state that $\sigma \leq 0.8^{+0.8}_{-0.6}$ (3-sigma
limits).  

In this context, it is interesting to note the recent claim of limits
$0.43 \lesssim \beta\Gamma \lesssim 1$ for the jet in Cyg X-1
\citep{gleissner:04}, which is part of the FP sample. Given the upper
limit on $\sigma$, this would place a 3-sigma upper
limit on $\beta\Gamma_{\rm mean} \leq 250$ and put Cyg X-1 at the low
end of $\beta\Gamma$ distribution.  In other words, if most of the
scatter in the distribution is indeed due to relativistic beaming,
then most of the jets in the sample should have faster velocities than
Cyg X-1.  The limit on $\beta\Gamma$ for Cyg X-1 is based on the lack
of correlations between radio and X-ray emission above a given
frequency.  If other XRB jet source are indeed significantly faster,
this should manifest itself correlations between radio and X-rays on
shorter timescales than in Cyg X-1, which can be tested
observationally.

The sample used to derive the FP contains some steep spectrum sources
and some sources without measured $\alpha_{\rm r}$.  As discussed in
\cite{merloni:03}, this can be an additional source of scatter.  In
order to assess the influence of the presence of steep spectrum
sources on the scatter about the fundamental plane and on the limits
we can place on the $\beta\Gamma$ distribution, we repeated the same
analaysis as above limited to sources that are known to have flat
radio spectra.  We find that the scatter is slightly reduced and that
the 1-sigma confidence contour moves downward to lower values of
$\sigma$, while the 2- and 3-sigma confidence contours
are expanded in all directions. This is because the number of sources
in the sample is reduced significantly, thus reducing the statistical
significance of the result.  The overall shape of the contours is not
changed, and the main conclusion that one can only place an upper
limit of $\sigma \leq 0.4^{+1.2}_{-0.4}$ from these
considerations remains.

\subsection{Beamed X-rays}
If we try to reproduce the scatter about the FP in a model where the
X-rays are produced in the jet at the same $\Gamma$ as the radio
(e.g., as synchrotron or synchrotron-self-Compton radiation), the
formalism changes: Assuming the X-ray and radio fluxes are emitted
with the same $\Gamma$ and the same viewing angle, and taking the
X-ray flux to follow a powerlaw of the form $F_{\rm x} \propto
\nu^{-\alpha_{\rm x}}$, the observed deviation of the radio flux from
the FP defined in eq.~(\ref{eq:delta}) is
\begin{eqnarray}
	\delta(P,\Gamma) = \frac{\Gamma^{0.6\alpha_{\rm x} - \alpha_{\rm r} -
0.4k}\left[\frac{1}{\left(1 + \beta P\right)^{k+\alpha_{\rm r}}}
+ \frac{1}{\left(1 - \beta P\right)^{k+\alpha_{\rm
r}}}\right]}{\left[\frac{1}{\left(1 + \beta P\right)^{k+\alpha_{\rm
x}}} + \frac{1}{\left(1 - \beta P\right)^{k+\alpha_{\rm
x}}}\right]^{0.6}}
\label{eq:jetbeaming}
\end{eqnarray}
For $k \sim 2$, $\alpha_{\rm r} \sim 0$, and $\alpha_{\rm x} \sim 1/2$
(typical for optically thin synchrotron emission), it turns out that
eq.~(\ref{eq:jetbeaming}) requires unrealistically large values of
$\Gamma$ to obtain the observed scatter about the FP, as plotted in
the right panel of Fig.~\ref{fig:chisquare}. The range in $\Gamma$
implied by the 1-sigma contours on $\beta\Gamma_{\rm mean}$ and
$\sigma$ would reach from $\Gamma$ of order unity to
$\Gamma \sim 10^5$ or higher.  Furthermore, in many sources the X-ray
spectra are steeper than $\alpha_{\rm x}=1/2$.  As can be seen from
eq.~(\ref{eq:jetbeaming}), the effectiveness of beaming to produce
scatter about the FP is reduced further when $\alpha_{\rm x}$ is
increased from 0.5 to 1 (in the latter case, values of
$\beta\Gamma_{\rm mean} \sim 10^8$ and $\sigma\sim
10$ are required to produce the observed amount of scatter).

Two possible conclusions arise from this result: If the X-rays are
produced by the jet, then either a) some other source of statistical
uncertainty must be present to dominate the observed scatter about the
FP, and/or b) the X-ray emission must arise from a region of the jet
that suffers less relativistic beaming.  Most jet acceleration models
actually accelerate the jet over several decades in distance to the
core.  The latter scenario would therefore be compatible with the
general notion that the optically thin X-ray synchrotron emission is
dominated by the innermost region of the jet, closest to the core,
while the optically thick radio emission stems from a region further
out that might have been accelerated to larger $\Gamma$.

Simple direct synchrotron models do present other challenges
\citep{heinz:04}.  More realistic scenarios include a combination of
synchrotron plus synchrotron-self-Compton and inverse Compton
scattering of disk radiation \citep{markoff:04}.  It is not clear
whether the X-ray emitting region in this scenario would be co-spatial
with the radio emitting region or not.  Certainly, however, the modest
amount of scatter in the XRB radio-X-ray relation and in the FP
relation cannot be used to argue in favor of a jet origin of the X-ray
- both disk X-rays and X-rays from the base of the jet can easily
produce the observed amount of scatter.

\subsection{Blazars and highly beamed sources}

As mentioned in \S\ref{sec:boosting}, in the absence of velocity
constraints on indicidual source (like those on Cyg X-1 used above),
the only way to obtain an upper limit on $\beta\Gamma$ from this
method is to observe the cutoff at high luminositis where the sources
fall into the beaming angle and no further amplification is possible.
However, in those sources the X-rays almost certainly contain a beamed
component from the jet, as observed in blazars and BL-Lacs.  Thus, the
source of the X-rays is possibly not the same as in the unbeamed
sources and the upper cutoff will not adequatly sample the maximum
$\Gamma$.  Furthermore, the sample used here was selected to exclude
blazars and BL-Lac objects (with the exception of 3C279) since they
are almost strongly selection biased and because the X-rays most
likelty come from a different source.  Thus we have specifically
eliminated the possibility to sample the upper cutoff even if it were
observable.

Following eq.~(\ref{eq:jetbeaming}), the effect of an additional,
strongly beamed X-ray component is to {\em reduce} the deviation from
the regular FP relation that would otherwise be measured for a large
positive beaming of the radio flux alone.  For a truly randomly
oriented, unbiased sample, the large majority of the sources will not
be stringly affected by this, because at high $\beta\Gamma$, a very
small fraction of sources falls into the beaming cone, while at low
$\beta\Gamma$, beaming is unimportant.  Since we cannot be sure that
the FP sample is free of bias, a note of caution is in order regarding
possible selection effects.  Still, because the conclusions reached in
this paper are not based on claims about the upper cutoff in the flux
distribution, the results should be robust even if the contribution
from highly beamed sources is not treated entirely self-consistently.

\section{conclusions}
\label{sec:conclusion}
We showed that the scatter in the radio-X-ray relation in XRBs and in
the ``fundamental plane'' relation in accreting black holes can be
used to constrain the width of the Lorentz factors distribution of the
jets in these sources.  It cannot be used to put an upper limit on the
mean Lorentz factor $\langle \Gamma \rangle$ of the jets in the
sample.  However, if all of the scatter is indeed due to relativistic
Doppler boosting, we show that a lower limit can be put on $\langle
\beta\Gamma \rangle$.  Both log-normal and log-flat distributions in
$\beta\Gamma$ fit the observed scatter well.  We show that, if the
X-rays are produced in the jet, they either have to originate in an
unbeamed portion of the jet (close to the base) or other sources of
scatter must dominate in the ``fundamental plane'' relation.
\vspace*{6pt}

\thanks{We would like to thank Rob Fender, Sera Markof, Mike Nowak,
    and the anonymous referee for helpful insights and
    discussions. Support for this work was provided by the National
    Aeronautics and Space Administration through Chandra Postdoctoral
    Fellowship Award Number PF3-40026 issued by the Chandra X-ray
    Observatory Center, which is operated by the Smithsonian
    Astrophysical Observatory for and on behalf of the National
    Aeronautics Space Administration under contract NAS8-39073.}

\end{document}